\documentclass[conference]{IEEEtran}
\IEEEoverridecommandlockouts
\usepackage{cite}
\usepackage{amsmath,amssymb,amsfonts}
\usepackage{algorithmic}
\usepackage{graphicx}
\usepackage{textcomp}
\usepackage{xcolor}
\usepackage{hyperref}
\usepackage{subfigure}

\usepackage[T1]{fontenc}
\usepackage{multirow}
\usepackage{algorithmic}
\usepackage{algorithm}
\usepackage{color}
\newcommand{\eat}[1]{}
\newcommand{\stitle}[1]{\vspace{1ex} \noindent{\emph{#1}}}

\def\BibTeX{{\rm B\kern-.05em{\sc i\kern-.025em b}\kern-.08em
    T\kern-.1667em\lower.7ex\hbox{E}\kern-.125emX}}
\begin{document}

\title{LR-CNN: Lightweight Row-Centric Convolutional Neural Network Training for Memory Reduction}

\author{
    \IEEEauthorblockN{Zhigang Wang}
    \IEEEauthorblockA{
        \textit{Ocean University of China}\\
        wangzhigang@ouc.edu.cn}
    \and

    \IEEEauthorblockN{Hangyu Yang}
    \IEEEauthorblockA{
        \textit{Ocean University of China}\\
        yanghangyu3961@stu.ouc.edu.cn}
    \and

    \IEEEauthorblockN{Ning Wang}
    \IEEEauthorblockA{
        \textit{Ocean University of China}\\
        wangning8687@ouc.edu.cn}
    \and

    \IEEEauthorblockN{Chuanfei Xu}
    \IEEEauthorblockA{
        \textit{Huawei Technologies Co., Ltd.}\\
        xuchuanfei@huawei.com}
    \and

    \IEEEauthorblockN{Jie Nie}
    \IEEEauthorblockA{
        \textit{Ocean University of China}\\
        niejie@ouc.edu.cn}
    \and

    \IEEEauthorblockN{Zhiqiang Wei}
    \IEEEauthorblockA{
        \textit{Ocean University of China}\\
        weizhiqiang@ouc.edu.cn}
    \and

    \IEEEauthorblockN{Yu Gu}
    \IEEEauthorblockA{
        \textit{Northeastern University}\\
        guyu@mail.neu.edu.cn}
    \and

    \IEEEauthorblockN{Ge Yu}
    \IEEEauthorblockA{
        \textit{Northeastern University}\\
        yuge@mail.neu.edu.cn}
}

\maketitle

\begin{abstract}
In the last decade, Convolutional Neural Network with a multi-layer architecture has advanced rapidly. However, training its complex network is very space-consuming, since a lot of intermediate data are preserved across layers, especially when processing high-dimension inputs with a big batch size. That poses great challenges to the limited memory capacity of current accelerators (e.g., GPUs). Existing efforts mitigate such bottleneck by external auxiliary solutions with additional hardware costs, and internal modifications with potential accuracy penalty. Differently, our analysis reveals that computations intra- and inter-layers exhibit the spatial-temporal weak dependency and even complete independency features. That inspires us to break the traditional layer-by-layer (column) dataflow rule. Now operations are novelly re-organized into rows throughout all convolution layers. This lightweight design allows a majority of intermediate data to be removed without any loss of accuracy. We particularly study the weak dependency between two consecutive rows. For the resulting skewed memory consumption, we give two solutions with different favorite scenarios. Evaluations on two representative networks confirm the effectiveness. We also validate that our middle dataflow optimization can be smoothly embraced by existing works for better memory reduction.
\end{abstract}


\section{Introduction}\label{section:1intro}
In recent years, Convolutional Neural Network (CNN) as one of the most successful deep learning technologies, has been widely used in various domains, especially for image understanding tasks. Till now, representative CNN models include VGG~\cite{vgg}, Inception~\cite{inceptionv4}, and ResNet~\cite{resnet}. All of them employ a multi-layer network architecture and are iteratively trained. The total input matrices (like images) are partitioned into batches and then used in a round-robin manner. Within an iteration, (1) a batch of inputs is firstly forwarded across layers to extract features and make the final prediction (i.e., Feed-Forward Propagation, {\bf FP}), (2) followed by loss calculation between prediction and ground-truth to measure how well the network is performing, and then (3) loss-based derivatives are calculated and propagated from the last layer through to the first, to refine parameters and their weights for loss reduction (i.e., Backward Propagation, {\bf BP}).

To address complex datasets and tasks, CNN is becoming deeper and larger, with hundreds of layers and kernel parameters. Its training thereby yields strong and ever-growing demands on memory resources. However, the de facto model size is fairly small (usually less than 500MB), because parameters as well as weights can be shared for a given input. By contrast, refining operations in BP depend on gradients, and the latter requires to preserve intermediate output-input (i.e., {\bf feature maps}) between any two consecutive FP layers. The volume of such data can be very large, especially when processing high-dimension matrices with a big batch size. Taking global climate analysis as an example, its Community Earth System Model can output millions of ultra-high-resolution images with up to 3,600$\times$2,400 pixels~\cite{ocean}. For the classic ResNet50 model with a small batch size 8, the accumulated feature maps roughly consume 120GB memory. That clearly exceeds the 80GB capacity of the most expensive accelerator Nvidia A100, leting alone other economic products like RTX3090 used in our experiments with only 24GB GPU memory.

To break the memory wall, researchers have devoted great efforts from aspects of external auxiliary and internal modification. The former typically includes distributing training onto multiple devices~\cite{reference:1,reference:3,reference:5}, and offloading data and even compute from GPU to CPU~\cite{reference:28,reference:30}. Both of them generate additional hardware costs and inevitably suffer from frequent data migration. The latter includes (1) checkpointing where only feature maps of partial layers are preserved and others are recomputed in BP~\cite{reference:34}, (2) compressing data by quantization and mixed precision~\cite{reference:13,reference:19}, and (3) pruning redundant and non-informative network configurations~\cite{reference:23,reference:24}. By our tests, checkpointing cannot work well as expected because of its built-in constraint; the other two usually have negative impacts on the model accuracy.

In general, existing works trade runtime latency, hardware investment, and accuracy loss for memory reduction. Although they achieve a significant improvement in memory scalability, there still is strong interest to further reduce the memory footprint, especially with lightweight penalty.

Our investigations find that the current CNN training assumes a complex many-to-many relationship between kernel parameters and matrices. That makes some sense because every small kernel matrix indeed needs to scan every large input image or the feature map matrix, to perform MUL-SUM convolution computations. This strong assumption means that computations cannot proceed to the next layer until all kernels within the current layer complete tasks. Such a layer-by-layer (column) dataflow design accumulates all feature maps, and hence can easily exhaust limited memory resources. But in fact, two MUL-SUMs are dependent only when matrix elements involved are partially overlapped. The dependency becomes weak and even disappears, when convolutions are far away from each other in the temporal-spatial scale. This insight offers an orthogonal direction for memory reduction, by breaking the column-centric dataflow constraint.

This paper firstly proposes a row-centric CNN training approach for memory reduction. Here kernel parameters are still organized in layers but the accompanied MUL-SUM computations for each kernel are divided into several rows and performed one by one. Towards this end, the original input or the intermediate feature map is also partitioned into row blocks. Now a kernel in layer-$i$ is normally slid but only elements in the scheduled row-block is available. The partial output feature map can be fed as a new row-bock input to convolutions in layer-($i$+1). This intra- and inter-layers dataflow optimizations allow the limited memory to be reused among different rows. The high peak memory usage is thereby dominated by the maximal consumption, instead of their sum. Note that the receptive field of a kernel might be covered by two consecutive row-blocks. Such information is commonly used in both rows to cope with the weak dependency, so as to guarantee the normal convergence. Besides, FP and BP employ different row granularities. Because feature maps between FP layers are removed, then a coarse-grained setting can reduce the coordination cost between dependent rows. By contrast, a fine-grained setting is given for BP so as to accommodate re-computed feature maps involved in the gradient calculation.

We also study two detailed yet important issues about row partitioning. One is the row granularity for the whole network. A large setting can achieve significant memory reduction, but the runtime latency increases due to low parallelism and high coordination costs. We state that it should be determined on demand in dedicated and multi-tenant environments. Another problem is that the memory footprint distribution among rows is usually skewed, due to the impact of convolution operations on the common receptive field. Moreover, such skewness dynamically changes as convolution proceeds. As a result, the resource allocated by the maximum requirement will be wasted, when a row with low demand on memory is processed.

We give two balancing solutions to crack this skewness nut. The first is two-phase sharing where the common part is exclusively computed within a row and then preserved in FP and BP phases, for being reused by the next row and gradient calculation. Data preservation inevitably interrupts normal computations but is insensitive to the compute power. It thereby especially works well on low-configured devices. Our second alternative is overlapping partitioning where the common receptive field causing skewness is divided and replicated into related rows to eliminate dependency. We perform an even division policy for better memory reduction. This design incurs redundant but easy-to-be-parallelized computations for overlapped data. Thus, it is particularly suitable for high-configured devices. Note that for both solutions, their row boundary and overlapping/preserving-size depend on the number of layers (depth). We deduce them from the last layer to the first. Besides, we make in-depth analysis between additional overlapping/preserving storage costs and memory savings. We formally give the upper bound of row granularities for the two solutions, to respectively guarantee normal training and minimize the overall memory footprint.

It is worth noting that we also inherit prominent designs in existing internal and external optimizations, respectively including abandoning cheap-to-be-recomputed data~\cite{reference:35,reference:36}, and checkpointing~\cite{reference:34}. Then more layers can be involved in row-centric update, to further boost benefits. That is especially important for deep networks.

The major contributions are summarized below.
\begin{itemize}
    \item Proposing a Lightweight Row-centric CNN training approach \emph{LR-CNN}, which re-organizes convolutions into rows in both FP and BP for memory reduction, without any loss of accuracy and additional hardware costs.
    \item Proposing two row partitioning solutions with very different designs. The accompanied analysis can help to gain optimal performance based on hardware configurations while insulating end-users from tedious low-level details.
    \item Performing extensive experimental studies on two representative CNN networks VGG-16 and ResNet-50. \emph{LR-CNN} yields up to 78\% improvement on memory reduction even with only GPU memory, compared against up-to-date competitors (w/ and w/O CPU memory).
\end{itemize}

The remainder of this paper is organized as follows. Sec.~\ref{section:2pre} provides necessary background introduction on CNN. Sec.~\ref{section:3row} highlights how to train CNN in our new row-centric manner. Sec.~\ref{section:4par} presents the detailed row partitioning implementations tailored to different environments. Sec.~\ref{section:5exp} reports evaluation results. Sec.~\ref{section:6related} overviews related works. Finally, Sec.~\ref{section:7con} concludes this paper and discusses future works.

\section{Preliminaries}\label{section:2pre}
This section briefly introduces CNN and then analyzes its space complexity, to show the challenge on memory resources.

\subsection{Convolutional Neural Network (CNN)}\label{section:2pre:cnn}
CNN is a classic deep learning architecture with tens and even hundreds of stacked convolutional and possible pooling layers, followed by several fully connected layers. Data go through all layers twice but in different directions, i.e., Forward Propagation ({\bf FP}) and Backward Propagation ({\bf BP}).

\stitle{FP.} Given an input sample tensor $X$, the core idea is to use a kernel parameter tensor $\theta$ across $L$ convolutional layers to respectively extract local features. The latter go through the subsequent full connection network to obtain a prediction tensor $Y$. Specifically, at the $l$-th convolutional layer, the output feature map $z^{l-1}$ from the previous layer is fed as input, so as to extract a new feature $z^{l}$ based on the local kernel $\theta^l$. Within this layer, perhaps there exists an additional internal pooling layer. It reduces the dimension of feature maps by replacing some values with a single max or average of them. Eq.~(\ref{equ:introfp}) mathematically show the convolution function. All $z^l$ and possible $\check{z}^l$ form the {\bf feature map data set} $\cup_{l=1}^{L}z^l$.

\begin{equation}
 z^{l}=\mathrm{Conv}(z^{l-1},\theta_{l})\\
 \label{equ:introfp}
\end{equation}

There also exist Activation and BatchNormalization functions in FP. However, the computation complexity is marginal due to simple principles. We can abandon the outputs like activations, and recompute them if necessary, as done in existing works~\cite{reference:35,reference:36}. We thereby exclude the related discussion from the scope of this paper.

\stitle{BP.} A specific loss function is designed to measure how well the network is trained, by calculating the error $\delta$ between predicted $Y$ and its ground truth $Y'$. $\delta$ is propagated in the reverse direction to deduce gradient $g$ based on feature map from the previous layer $z^{l-1}$. Then we can refine kernel $\theta$ by $g$. Eq.~(\ref{equ:introbp}) shows the process at the $l$-th layer.

\begin{equation}
 g^l = \mathrm{Gradient}(\delta^l, z^{l-1})
 \label{equ:introbp}
\end{equation}

\begin{figure*}[htbp]
\centering
\includegraphics[width=0.65\linewidth]{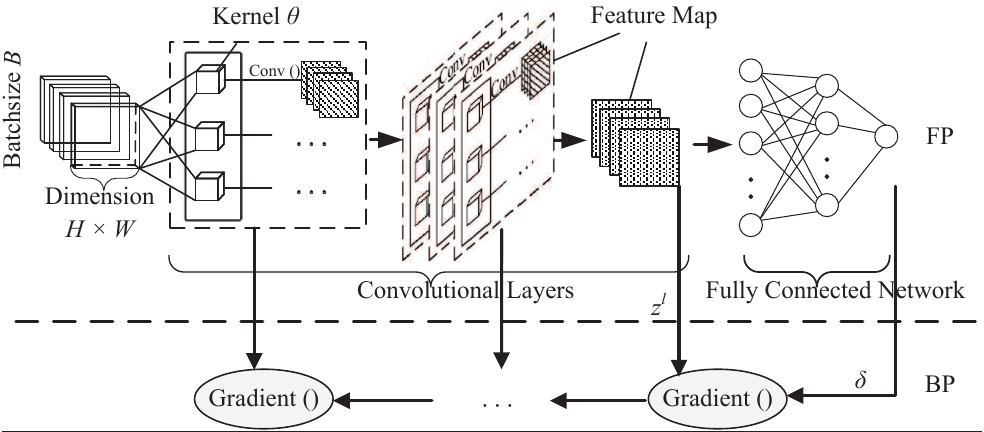}
\caption{Traditional CNN}
\label{figure:traditional_CNN}
\end{figure*}

As shown in Fig.~\ref{figure:traditional_CNN}, CNN is typically trained in a mini-batch manner where $X$ is divided into several batches and then fed into the network in a round-robin manner. Multiple rounds/epochs are required to refine $\theta$, until the overall loss converges. Each batch corresponds to an iteration consisting of FP and BP. The mainstream deep learning platforms like PyTorch build a dependency graph in FP to guide the gradient calculation in BP.

\subsection{Space Complexity of Training CNN}\label{section:2pre:space}
Sec.~\ref{section:2pre:space} tells us that there exist four kinds of data, kernel $\theta$, gradient $g$, error/loss $\delta$, and feature map. Among them, $\theta$ has the same data volume with $g$, which is usually small since CNN enables sharing parameters. Taking ResNet-50 as an example, it roughly consumes 200MB memory in PyTorch by our test. $\delta$ is also space-negligible, because it immediately disappears once $g^l$ is computed at the specific $l$-th layer, instead of being accumulated. In conclusion, the memory footprint is mainly dominated by feature map. Below, we give its space complexity.

The feature map data set is rebuilt at every iteration. Given a specific iteration, assume that the related input batch consists of $B$ matrices with $H\!\times\!W$ dimensions. Eq.~(\ref{equ:introspace}) shows the global size of feature maps accumulated across $L$ layers. $\varrho^{l}$ is the size specific to the $l$-th layer, where $H^{l}$ and $W^{l}$ are the dimensions of $z^l$, and $C^l$ stands for the number of channels/kernels $\theta^l$.

\begin{equation}
 \Omega = \sum_{l=1}^{L} \varrho^{l} = \sum_{l=1}^{L}(B\cdot H^{l}\cdot W^{l}\cdot C^{l})
 \label{equ:introspace}
\end{equation}

Eq.~(\ref{equ:introspace}) reveals that the feature map size linearly grows with the increase of batch size $B$ and dimensions $H\!\times\!W$. In turn, given the limited memory capacity, it is difficult for data scientist to adjust the batch size to a desired value, especially when large input matrices are fed, like high-resolution images. Since there exists the semantic constraint on $B$ for model quality, the only space opening for us is to tune $H$ and $W$, namely, the row-centric training mechanism presented in Secs.~\ref{section:3row} and \ref{section:4par}.

\section{LR-CNN: Lightweight Row-centric \\CNN Training}\label{section:3row}
This section introduces our row-centric CNN training framework. Our solution is lightweight because no additional hardware is required. Below we first give the overview design, followed by a convergence discussion, and finally introduce the constraint of row granularities in FP and BP.

\subsection{Overview Design}\label{section:3row:partition}

Before introducing our overview design, we first analyze the computation dependency among core convolution operations in CNN. That is the foundation of our proposals.

\stitle{Strong or Weak Dependency?}
At the $l$-th layer, a convolution slides every kernel in $\theta^l$ from left-to-right and top-to-bottom, along every feature map matrix in $z^{l-1}$ from the previous layer. This is conventionally modeled as a complex many-to-many relationship. Such {\bf strong dependency} enforces a layer-by-layer computation manner, and hence, feature maps across layers are accumulated. However, at a specific time instance, in fact only the region covered by the kernel is involved in converting elements within it into a single value, via MUL-SUM computations. This region is also called as Receptive Field ({\bf RF}). Data out of RF are clearly temporally independent, even though they reside in the same layer. Further, a MUL-SUM operation at the ($l\!+\!1$)-th layer can be immediately launched once elements in RF at the $l$-th layer are ready, yielding a spatial independency between layers. Such {\bf spatial-temporal independency} motivates us to partially perform convolutions throughout all convolutional layers and the release related feature map data immediately, to reduce memory footprint.

\stitle{Row-centric Convolution Scheduling.}
Inspired by the independency analysis, we re-organize all convolutions into rows throughput all convolutional layers. Let $N$ be the number of rows. Every input matrix $\chi\!\in\!X$ as well as the intermediate feature map $z^{l}$ is accordingly performed an end-to-end split, along the height dimension. Then $\chi\!=\!\sum_{r=1}^{N}\!\oplus\!\chi_{r}$ and $z^{l}\!=\!\sum_{r=1}^{N}\!\oplus\!z_{r}^{l}$, where $\oplus$ indicates the concatenation of submatrices. For the error tensor $\delta$, we also have $\delta^l\!=\!\sum_{r=1}^{N}\!\oplus\!\delta_{r}^{l}$. Eqs.~(\ref{equ:rcnnfp}) and (\ref{equ:rcnnbp}) respectively show how to extract feature map in FP and compute gradient in BP. Given the $i$-th row $R_i$, in FP, Eq.~(\ref{equ:rcnnfp}) is run layer-by-layer. Once the $l$-th layer is passed, we can immediately release the corresponding input feature map $z_i^{l-1}$, since it is no long required by other rows $R_j$ ($j\!\neq\!i$) and subsequent layers within $R_i$. That significantly reduces the memory footprint. For BP, note that now the required feature map $z_{i}^{l-1}$ is not available. Thus, it must recompute $z_{i}^{*}$ across all layers from the first to the last, and then run Eq.~(\ref{equ:rcnnbp}) in the reverse direction to compute gradients. Here, we still need to preserve feature maps across layers but only a subset for each. The memory footprint is still reduced.

\begin{equation}
 z_{i}^{l}=\mathrm{Conv}(z_{i}^{l-1},\theta_{l})\\
 \label{equ:rcnnfp}
\end{equation}

\begin{equation}
 g_{i}^l = \mathrm{Gradient}(\delta_{i}^l, z_{i}^{l-1})
 \label{equ:rcnnbp}
\end{equation}

Our row partitioning cannot cope with fully connected layers since the many-to-many connections among neurons exhibit strong dependency. Thus, the output of the last convolutional layer $z_{i}^{L}$ will not be released. Instead, after all rows are scheduled, we concatenate their outputs and then feed the final result $z^{L}$ into the full connection network. Thanks to multiple convolution and possible pooling operations, $z^{L}$ at the final convolutional layer is usually small. We can easily accommodate it. Another problem is that a kernel needs to seamlessly scan all elements in a feature map. Some RFs then inevitably cover two consecutive rows. In this case, the convolution computation should be jointly completed by the two rows. We will give a detailed analysis for such {\bf weak dependency} in Sec.~\ref{section:3row:forward}.

\begin{figure}[htbp]
\centering
\includegraphics[width=0.9\linewidth]{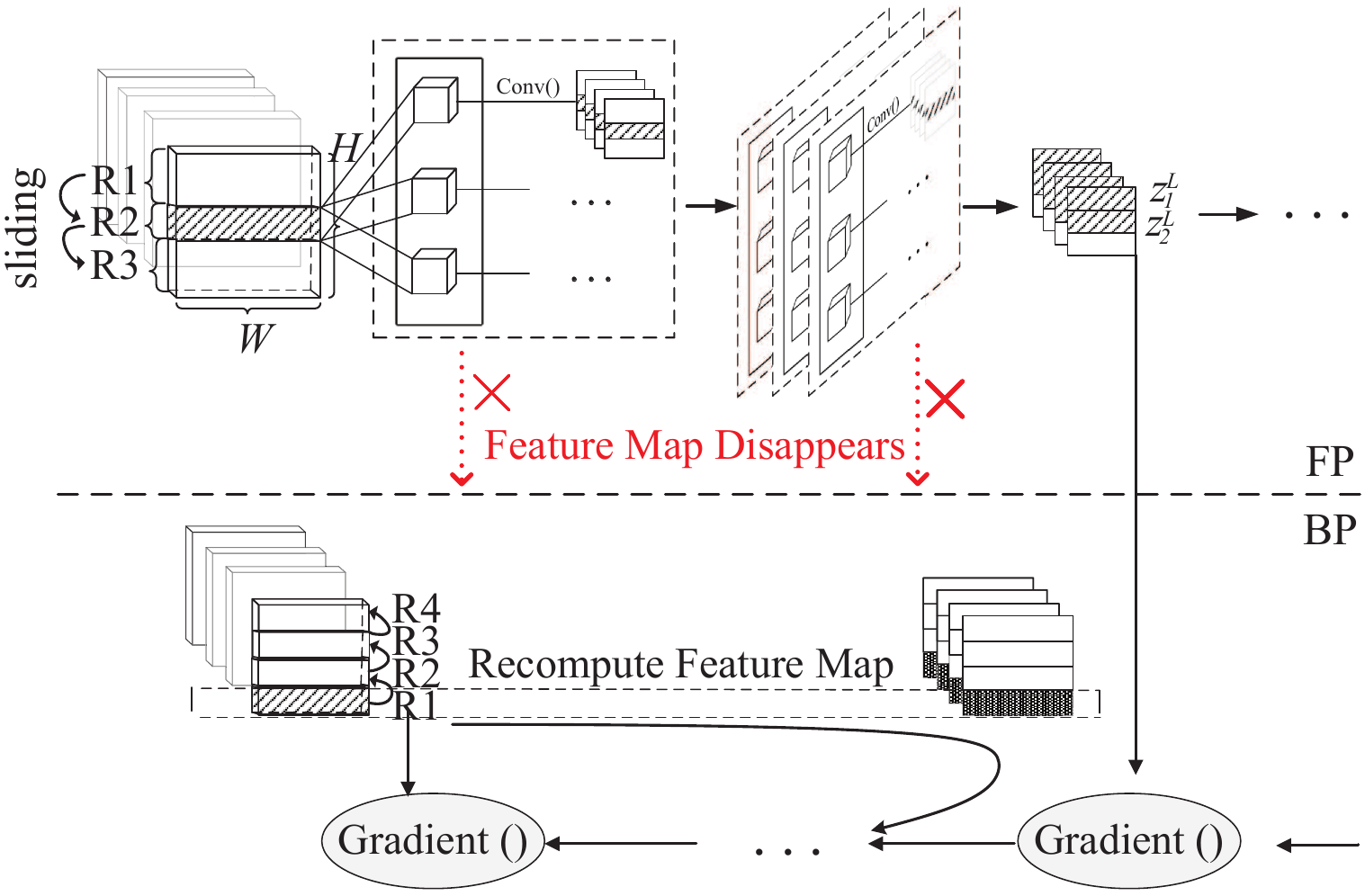}
\caption{Row-centric CNN}
\label{figure:LRCNN}
\end{figure}

\stitle{An Example.}
Fig.~\ref{figure:LRCNN} gives a concrete example with $N\!=\!3$, to demonstrate how our row-centric mechanism works. In FP, we schedule convolutions row-by-row. Assume that $R_i$ is completed, all related feature maps are released but the final outputs $z_{1}^{L}$ and $z_{2}^{L}$ are preserved and concatenated. Once $R_{3}$ is completed, we can get the correct and full $z^{L}$ like traditional column-centric computations, and then feed it into the full connection network for prediction. Afterwards, in BP, we recompute dropped feature maps from the bottom to the top, based on the dependency graph built in FP. Here we split matrices into more parts ($4\!>\!3$). Otherwise, a row in BP will consume more memory resources than that in FP, and then the peak consumption increases. This is because the former needs to accumulate sub-feature-map $z_i^{l}$ while the latter does not. The row number constraint between FP and BP is analyzed later in Sec.~\ref{section:3row:backward}.

Algorithm~\ref{algorithm:Row-centric_CNN_Training} finally outlines the process of our row-centric design within an epoch.
Each sample $\chi$ undergoes FP and BP. It is row-wise partitioned (Line~5), and then goes through $N$ FP layers. Feature maps between any two consecutive layers are released after computations (Lines~6-10). But the total $N$ rows of outputs at the final convolutional layer are concatenated to recover the original maps (Line~11), for subsequent full connection network (Lines~12-14). Afterwards, BP for each row in the convolutional layer is carried out to obtain the gradient $g$ (Lines~15-23). Once all gradients are ready, parameters $\theta$ are updated (Line~24). Note that it is necessary for BP to recompute feature maps released in FP (Line~17).
\begin{algorithm}
\caption{Row-centric CNN Training}
\label{algorithm:Row-centric_CNN_Training}
\begin{algorithmic}[1]
\STATE Initialize convolutional kernel $\theta$
\FOR {$t=1,...,Epochs$}
    \STATE// \textit{Iterations}
    \FOR {$\chi \in  X$}
        \STATE Split $\chi $ to $ \chi _{1}, \chi _{2}, \ldots, \chi _{N} $
        \FOR{$r=1,...,N$}
            \STATE ${z}_{r}^{0}\leftarrow \chi _{r}$
            \FOR{$l=1,...,L$}
                \STATE ${z}_{r}^{l} \leftarrow Conv({z}_{r}^{l-1},\theta^{l})$, delete ${z}_{r}^{l-1}$
            \ENDFOR
            \STATE $Z^{L} \leftarrow \sum_{r=1}^{N}\!\oplus z_{r}^{L}$
            \STATE $y'^{(i)} \leftarrow  Fully\_connected\_FP(z^{L})$
            \STATE $\delta  \leftarrow Loss(Y'^{(i)}, Y^{(i)})$
            \STATE $\delta^{L} \leftarrow Fully\_connectd\_BP(\delta )$
            \STATE split  $\delta^{L}$ to  $ \theta_{1}^{L}, \theta_{2}^{L},, \ldots, \theta_{N}^{L}, $
            \FOR{$r=N,...,1$}
                \STATE Recompute to get all $z_{r}^{l}$ layer by layer, $\forall l \in [1,L-1]$
               \FOR{$l=L,...,1$}
                    \STATE $\delta_{r}^{l-1} \leftarrow Get\_error(\delta_{r}^{l},\theta^{l})$
                    \STATE $g^{l} += Gradient(\delta_{r}^{l},z_{r}^{l-1})$
                \ENDFOR
            \ENDFOR
        \ENDFOR
        \STATE  $\forall l \in [1,L],\theta \leftarrow update(\theta^{l}, g^{l}), g^{l} = 0$, meanwhile updating fully connected layer weights
    \ENDFOR
\ENDFOR
\end{algorithmic}
\end{algorithm}

\subsection{Convergence Analysis}\label{section:3row:forward}
This section analyzes the negative impact of row partitioning on convergence. Our analysis reveals that there exist two problems, one is feature loss and another is feature redundancy. Both are related to the weak dependency. Below, we respectively show why they happen using concrete examples, and finally give the mathematical conclusion.

\begin{figure}[htbp]
  \centering
  \subfigure[Feature loss]{\includegraphics[width=\linewidth]{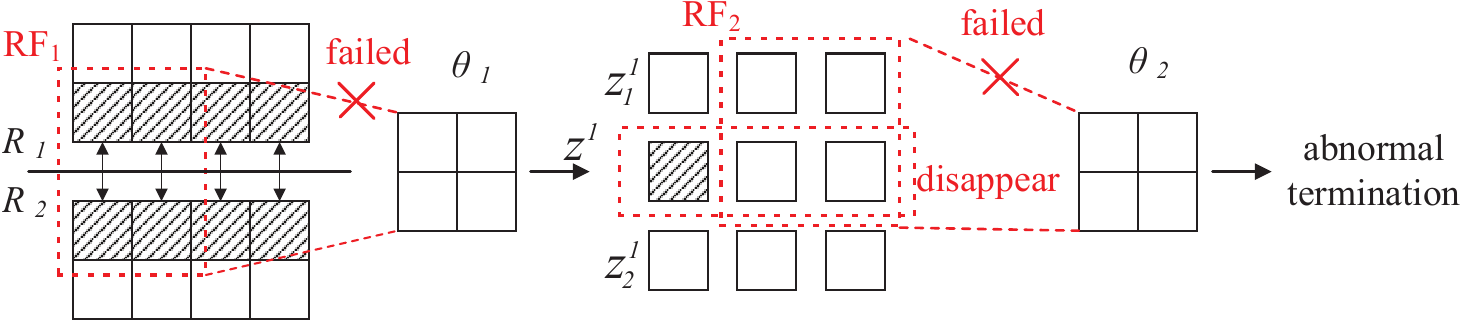}\label{fig:feture_loss}}
  \subfigure[Padding redundancy]{\includegraphics[width=\linewidth]{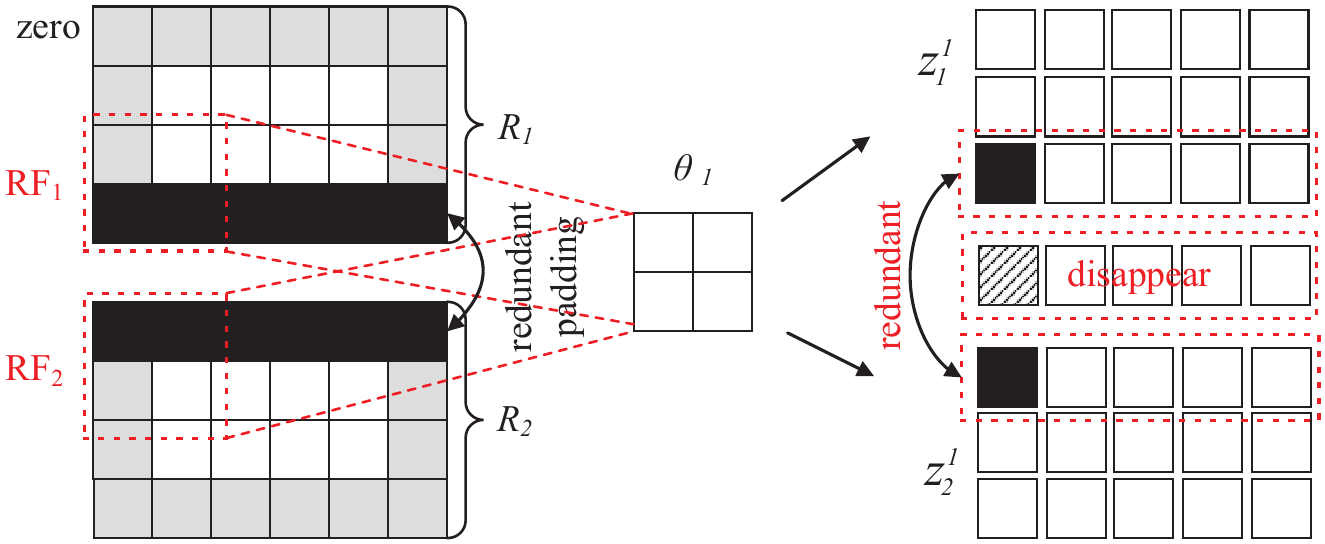}\label{fig:excess_padding}}
  \caption{Variation of feature map}
  \label{figure:feature_map_size_changing}
\end{figure}

\stitle{Feature Loss.}
Recall that a normal convolution computation will slide RF from left-to-right and top-to-bottom with stride $s$. Since we partition an underlying input matrix along the height dimension, all columns can be preserved in a row $R_i$. That naturally guarantees the left-to-right data scan over columns. However, the top-to-bottom sliding operation will be interrupted when some rows required by the RF are not included in the currently scheduled $R_i$. Fig.~\ref{fig:feture_loss} shows such weak dependency constraint. The input matrix with $H\!\times\!W\!=\!4\!\times\!4$ is partitioned into two rows $R_1$ and $R_2$. The two kernels have the same size with dimension $k\!=\!2$. Then the RF size is $4\!\times\!4$. Assuming the stride $s\!=\!1$, we can normally run convolutions in $R_1$ for the first left-to-right scan and obtain three output values. In the second scan, the bottom boundary row of $R_1$ and the top boundary row of $R_2$ should be convolved together in \emph{RF}$_1$. However, $R_1$ and $R_2$ are scheduled at different times, that cuts off such connection. The top boundary of $R_2$ is thereby not available, yielding a convolution failure. As a result, $R_1$ as well as $R_2$ only outputs a partial feature map with $1\!\times\!3$. The concatenated $z^{1}$ has $2\!\times\!3$ dimensions, instead of normal $3\!\times\!3$. The disappeared row yields a cascading failure on \emph{RF}$_2$. In fact, the convolution is terminated abnormally since the kernel $\theta_2$ is bigger than $z_1^1$. We call this phenomenon as feature loss, which clearly disturbs the normal training process and destroys the convergence.

\stitle{Padding Redundancy.}
With the sliding operation of a kernel, the boundary data of the input/feature map is involved in convolutions with low frequency, leading to information loss. Researchers thereby extend and pad outer boundaries with zeros, as shown in Fig.~\ref{fig:excess_padding}. However, after row partitioning, some inner elements now become boundary data and hence are also surrounded by padding zeros. Such inner elements are then convoluted more times than the normal column-centric version. Worse, this redundancy is equivalent to changing an original vector $(a, b)$ into $(a, 0, 0, b)$. The added 0 values directly break some connections between $a$ and $b$. Both factors yield an uncontrolled disturbance on convergence. As shown in Fig.~\ref{fig:excess_padding}, the redundant padding generates two additional RF, \emph{RF}$_1$ and \emph{RF}$_2$. They finally output redundant values in $z_1^1$ and $z_2^1$. And the feature loss phenomenon still exists.

\stitle{Conclusion and Solution.}
Given the input feature map $z^l$ with dimensions [$H$, $W$], under traditional column-centric update, the output feature map $z^{l+1}$ after a convolution should have dimensions $\left [\left \lfloor \frac{H+2p-k}{s}-1  \right \rfloor, \left \lfloor \frac{W+2p-k}{s}-1  \right \rfloor \right ]$. Here $p$ is the padding size. However, under our row-centric design, the $i$-th part $ z_{i}^{l} $ with [$\frac{H}{N}$,$ W$] yields an output $ z_{i}^{l+1} $ with $\left[ \left \lfloor \frac{\frac{H}{N}+2p-k}{s}-1  \right \rfloor, \left \lfloor \frac{W+2p-k}{s}-1  \right \rfloor \right]$ . And the final concatenated output has the size of $ \left[ \left \lfloor \frac{H+2Np-Nk}{s}-1  \right \rfloor, \left \lfloor \frac{W+2p-k}{s}-1  \right \rfloor \right]$. That is different from the normal size achieved under column-centric update.

When $p\!=\!0$, our height dimension size is always less than the normal value, indicating that some features are lost. Sec.~\ref{section:4opt:overlap} and Sec.~\ref{section:4opt:sharing} will give two different solutions by overlapping and two-phase methods.
When $p\!>\!0$, the comparison result is non-deterministic. We can even observe the equality if redundant elements rightly compensate disappeared ones, but the negative impact still exists because the numerical values are not correct. Our solution is to perform semi-closed padding where the new boundary generated by row partitioning will be ignored during padding. Together with techniques in Sec.~\ref{section:4opt:overlap} and Sec.~\ref{section:4opt:sharing}, we can guarantee convergence.

\subsection{Customizing Row Granularity in FP and BP}\label{section:3row:backward}
Because an iteration consists of sequentially executed FP and BP, the overall peak memory usage is dominated by the higher one of the two phases. An ideal scenario is that FP and BP have the same memory usage, without any waste of resources. However, for both of them, the specific usage heavily depends on the row number $N$. To achieve our goal, below we first give their space complexities. For simplicity, we assume that the feature map matrix is evenly partitioned into $N$ rows.

For FP, as described in Sec.~\ref{section:3row:partition}, all feature maps are finally concatenated at the $L$-th layer to prepare the input for the full connection network, yielding a $\varrho^{L}$ cost, the same with that in column-centric update (see Eq.~(\ref{equ:introspace})). But before that, any feature map $z_{i}^{l}$ is immediately released when the convolution at the ($l\!+\!1$)-th layer is completed. Since the number of kernels and the specific kernel size vary with layers, the memory usage during the previous $(L\!-\!1)$ layers is dominated by the maximum, i.e., $\mathrm{max}_{l\in[1, L-1]}\{\varrho_{i}^{l}\}$, where $\varrho_{i}^{l}$ represents the space complexity caused by $z_{i}^{l}$, as shown in Eq.~(\ref{equ:rowfeaturemap}).
\begin{equation}
  \varrho _{i}^{l} = B\cdot H_{i}^{l}\cdot W^{l}\cdot C^{l} = B\cdot \frac{H^{l}}{N}\cdot W^{l}\cdot C^{l}
  \label{equ:rowfeaturemap}
\end{equation}
Then the total cost of FP, $\Omega_{\mathrm{FP}}$, is given by Eq.~(\ref{equ:fpspaceest}).

\begin{equation}
  \Omega_{\mathrm{FP}}(N) = \mathrm{max}_{l\in[1, L-1]}\{\frac{\varrho^{l}}{N}\} + \varrho^{L}
  \label{equ:fpspaceest}
\end{equation}

In BP, as required by Eq.~(\ref{equ:rcnnbp}), we need to recompute from the input to provide feature map $z_{i}^{l-1}$, to infer gradient $g_i^l$ at the $l$-th layer. If we still forget the intermediate results $z_{i}^{x}$, $x\!\in\![1, l\!-\!1]$, the recomputation must be repeatedly run, yielding prohibitively expensive runtime costs. We there cache these data but at expense of memory consumption. Similarly, Eq.~(\ref{equ:bpspaceest}) gives the space complexity of BP.

\begin{equation}
  \Omega_{\mathrm{BP}}(N) = \sum_{l=1}^{L-1}\frac{\varrho^{l}}{N} + \varrho^{L}
  \label{equ:bpspaceest}
\end{equation}

We can easily find that if FP and BP employ the same row granularity $N$, there exists $\Omega_{\mathrm{BP}}\!>\!\Omega_{\mathrm{FP}}$ due to caching feature maps. The peak memory usage of an iteration $\Omega\!=\!\mathrm{max}\{\Omega_{\mathrm{FP}}, \Omega_{\mathrm{BP}}\}\!=\!\Omega_{\mathrm{BP}}$. There will be a considerable waste of resources for FP. Thus, we need to customize their granularities $N_{\mathrm{FP}}$ and $N_{\mathrm{BP}}$ individually. We have two principles for that.
\begin{itemize}
    \item The total space complexity cannot exceed the available memory capacity $\mathcal{M}$, to avoid ``out-of-memory'' failures.
    \item A small setting is preferred for both $N_{\mathrm{FP}}$ and $N_{\mathrm{BP}}$, so as to fully utilize the parallel compute power.
\end{itemize}

Let $\xi$ present the memory consumption of all other data, typically including kernels and gradients. Then $N_{\mathrm{FP}}$ and $N_{\mathrm{BP}}$ are inferred by respectively enforcing inequations (\ref{equ:idlefpn}) and (\ref{equ:idlebpn}).

\begin{equation}
  \mathrm{min}(N_{\mathrm{FP}})\quad s.t. \quad \Omega_{\mathrm{FP}}(N_{\mathrm{FP}}) + \xi < \mathcal{M}
  \label{equ:idlefpn}
\end{equation}

\begin{equation}
  \mathrm{min}(N_{\mathrm{BP}})\quad s.t. \quad \Omega_{\mathrm{BP}}(N_{\mathrm{BP}}) + \xi < \mathcal{M}
  \label{equ:idlebpn}
\end{equation}

In particular, we have $N\!=\!N_{\mathrm{BP}}$ because $N_{\mathrm{BP}}$ is always smaller than $N_{\mathrm{FP}}$.

\section{Row Partitioning with Complex-objectives}\label{section:4par}
We finally present detailed implementations of row partitioning. We particularly study how to cope with the weak dependency between two consecutive rows.

\subsection{Two-phase Partitioning by Sharing Inter-row Computations}\label{section:4opt:sharing}
Sec.~\ref{section:3row:backward} lists two goals of row partitioning, i.e., we should minimize the granularity $N$ to fully utilize the strong parallel compute power, while guaranteeing memory safety based on the real-world constraint. In Eqs.~(\ref{equ:idlefpn}) and (\ref{equ:idlebpn}) give the answer but with a strong even-partitioning assumption. However, it cannot hold true due to the weak dependency between two consecutive rows.

\begin{figure}[htbp]
  \centering
  \subfigure[Skewed initial partitioning]{\includegraphics[width=\linewidth]{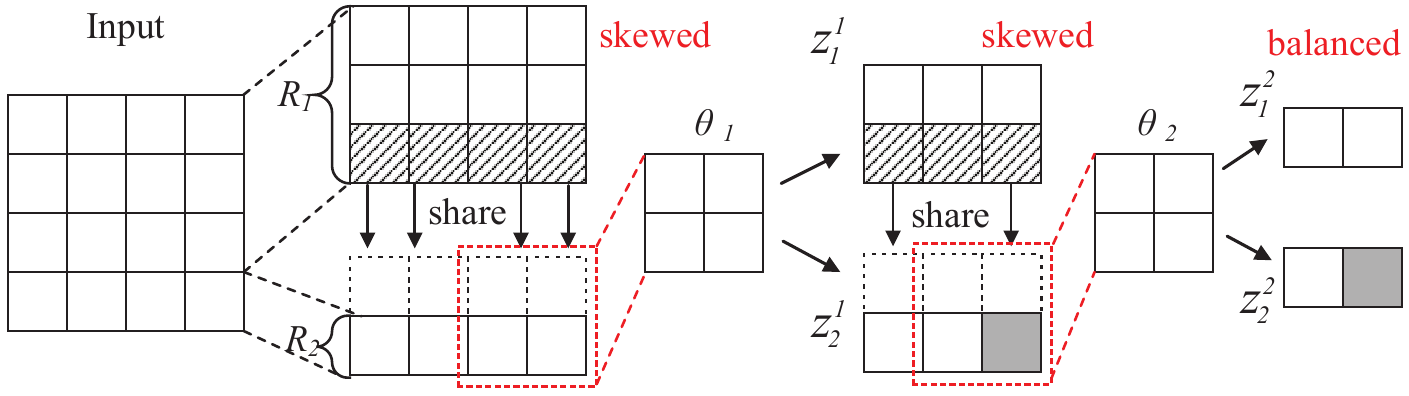}\label{figure:Two-Phase:1}}
  \subfigure[Skewed intermediate partitioning]{\includegraphics[width=0.78\linewidth]{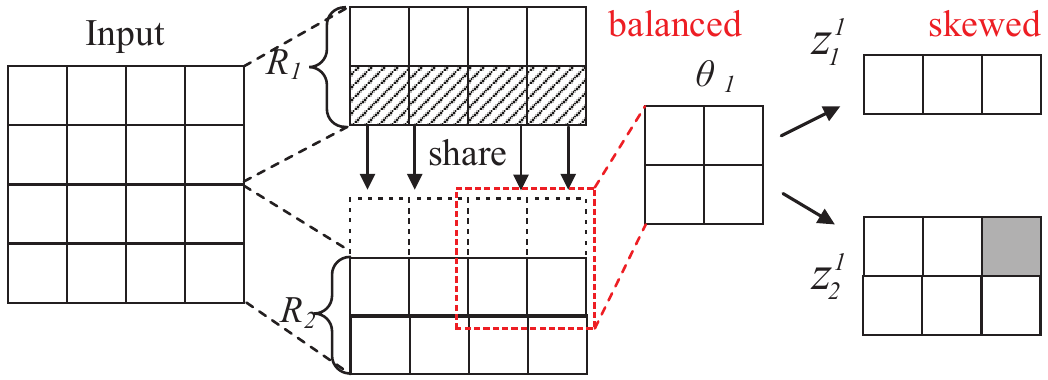}\label{figure:Two-Phase:2}}
  \caption{Illustration of skewness in \emph{2PS}}
  \label{figure:Two-Phase}
\end{figure}

\stitle{Problem Analysis.}
The weak dependency essentially requires data partitioned into two rows $R_i$ and $R_{i+1}$ are jointly convoluted in some RF. But $R_i$ and $R_{i+1}$ are scheduled at different times by our design. To cope with this conflict, a straightforward solution is to cache partial results computed in $R_i$ but also required by $R_{i+1}$. These results are reused/shared when $R_{i+1}$ is scheduled in both FP and BP phases. We thereby call it as {\bf Two-Phase Sharing}, or \emph{2PS} for short. However, it brings a heavily skewed memory consumption distribution across rows, which is difficult for balance again. As shown in Fig.~\ref{figure:Two-Phase:1}, we make a skewed initial partitioning for the input matrix where $R_{1}$ has dimensions [3,4] and $R_{2}$ has [1,4]. Benefitting from sharing data from $R_1$,  $R_2$ can normally perform convolution at the $1$st layer. But the output $z_1^1$ and $z_2^1$ still have skewed dimensions. After another convolution, we achieve the balance goal for $z_1^2$ and $z_2^2$. Fig.~\ref{figure:Two-Phase:1} shows another case where $R_1$ and $R_2$ are initially balanced. But the number of dimensions of $R_1$ quickly decreases, yielding a skewed distribution for $z_1^1$ and $z_2^1$.

\stitle{Greedy Row Partitioning.} Balanced data distribution of course can minimize the peak memory consumption and then improve the resource utilization. However, Fig.~\ref{figure:Two-Phase} reveals that no matter how to partition rows, the skewness always exists at different layers, due to multiple convolution operations. It is difficult to consistently guarantee balance. We thereby move our focus from the indirect balance goal to the direct utilization goal.

By Fig.~\ref{figure:Two-Phase}, another observation we can make is that the first  row $R_1$ reduces its dimension size with a unique damping factor. By contrast, others share the same but another smaller factor, if and only if $R_{i}$ with $i\!>\!1$ can always share data from $R_{i-1}$. This is because $R_1$ cannot acquire any extra data while $R_{i}$ can do that to slow down the size reduction. Inspired by this, we devise a greedy partitioning policy. We first focus on $R_1$. We infer its initial partitioning height $H_{1}^{0}$ at the input matrix in a reverse direction, i.e., from the last convolutional layer $H_1^L$ to $H_{1}^{0}$, by Eq.~(\ref{equ:2psditui}). That guarantees that $R_2$ can always share data from $R_1$, since the latter perform convolution operations through all convolutional layers. Then we can infer $\varrho _{i}^{l}$ by Eq.~(\ref{equ:rowfeaturemap}). In particular, we compute $H_1^0$ and $N_{\mathrm{BP}}$ under BP, since BP consumes more memory resources than FP. Now we need to additionally pay for the cached sharing sub-feature-map with size $(k^l\!-\!s^l)\!\cdot\!W^l$, at every convolutional layer except the last one, and at every row except the first one $R_1$. Generally, we should maximize $H_1^L$ fully and safely utilize memory resources, while minimizing $N_{\mathrm{BP}}$ for better parallelism, reducing the temporal-spatial cost of caching sharing data. As shown in Eq.~(\ref{equ:2psbpfirstheight}) we enumerate possible $H_{1}^{L}$ and $N_{\mathrm{BP}}$ to satisfy our constraints.

\begin{equation}
  H_{1}^{l} = (H_{1}^{l+1}  - 1) \cdot s^l + k^l - p^l
  \label{equ:2psditui}
\end{equation}

\begin{align}
&\max{H_{1}^{L}}\quad \mathrm{and}\quad\min{N_{\mathrm{BP}}} \quad s.t.\notag\\
&\sum_{l=1}^{L-1}\varrho_{1}^{l} + B(N_{\mathrm{BP}}-1)\sum_{i=1}^{L-1}(k^l\!-\!s^l)\!\cdot\!W^l\!\cdot C^{l} + \varrho ^{L} + \xi < \mathcal{M}
  \label{equ:2psbpfirstheight}
\end{align}

\stitle{Impact of} $N\!=\!N_{\mathrm{BP}}$.
Since $N\!=\!N_{\mathrm{BP}}$, we use $N$ in the following analysis.
Furthermore, Eqs.~(\ref{equation:row_2_to_N-1}) and (\ref{equation:last_row}) respectively show the computation of heights for some middle row $R_r$ and the last row $R_N$. They are different from $R_1$ since both $R_r$ and $R_N$ will share data from the previous row, and $R_N$ possibly has additional padding values.
\begin{equation}
 H_{r}^l = (H_{r}^{l+1} - 1) \cdot s^l + s^l \label{equation:row_2_to_N-1}
\end{equation}
\begin{equation}
 H_{N}^l = (H_{N}^{l+1} - 1) \cdot s^l + s^l - p^l\label{equation:last_row}
\end{equation}

By Eqs.~\ref{equ:2psditui}, \ref{equation:row_2_to_N-1} and \ref{equation:last_row}, we found that a large layer number $L$ will increase the times of the recursive computation. The height difference among $R_1$, $R_r$, and $R_N$ significantly increases, leading to a large memory consumption gap. The overall peak consumption is clearly dominated by the maximum value. As a result, further reducing the consumption of others does not make sense. Then increasing $N$ cannot work well. Since the underlying reason is the over-large $L$, our solution is to integrate checkpointing into our \emph{2PS}, to truncate layers. Then \emph{2PS} can work between two consecutive checkpointing locations with a relatively small layer number. Nevertheless, the cost of caching sharing data is still proportional to $N$. Once it is greater than the maximum cost of $R_1$, $R_r$, and $R_N$, i.e., $(N-1)(k^l\!-\!s^l)\!>\!\max\{H_1^l, H_r^l, H_N^l\}$, the overall memory usage increases.

\subsection{Overlapping Partitioning for Independent Computations}\label{section:4opt:overlap}
Although \emph{2PS} can improve memory scalability, it frequently interrupts normal convolution computations for caching sharing data. More specifically, it needs to extract a submatrix from $z_i^{l}$ and concatenate it with $z_{i+1}^{l}$ to form a new feature map. The total number of such operations is up to $(N\!-\!1)\sum_{i=1}^{L-1}$. These interruptions heavily decrease the throughput of modern accelerators with massive parallel threads. The proportionally increased memory allocation and collection operations are also time-consuming.

To overcome the underutilization of compute power, now we propose another alternative, i.e., the overlapping approach. It enforces two consecutive rows to collaboratively process dependent data. Its primary objective is to achieve an even row partitioning results and fully utilize the ever-growing power.

\begin{figure}[htbp]
\centering
\includegraphics[width=\linewidth]{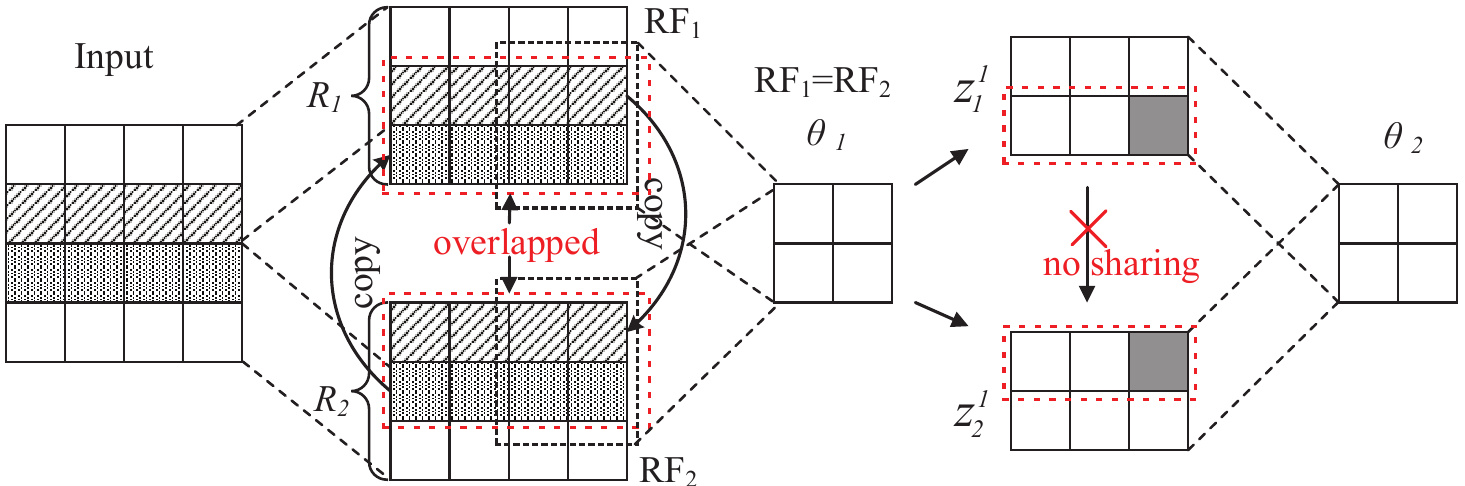}
\caption{Illustration of overlapping partitioning}
\label{figure:OverLapping}
\end{figure}

\stitle{Overlapping Partitioning.}
Our main idea is to replicate dependent data across different rows. Such overlapping partitioning enables each row to complete the convolution operations individually, without any coordination, like extracting and concatenating matrices. More importantly, each row has equal-sized dimensions and enjoys the same damping factor. That completely eliminates any possible runtime skewness, which facilitates an even partitioning so as to fully utilize memory resources. Fig.~\ref{figure:OverLapping} gives an example using an input matrix with $4\!\times\!4$ dimensions. Due to the $2\!\times\!2$ kernel, the two rows $R_1$ and $R_2$ require additional data from each other. They pull such data before training to avoid runtime interruption. These redundant data help $R_1$ and $R_2$ to complete convolution operations respectively in \emph{RF}$_1$ and \emph{RF}$_2$. In fact, elements involved in the two receptive fields are the same, leading to redundant computations. However, such matrix computations can be easily parallelized in currently mainstream deep learning platforms like PyTorch. The runtime latency is acceptable especially when enough computer power is provided.

Note that redundant computation will make skewed contributions to parameter refinement. We need to eliminate such negative impact. In particular, during the computation of the row containing $z_{r+1}^{l}$, $z_{r+1}^{l}[:b^{l}]$ is redundantly computed due to data replication. During BP, for gradients of weights from redundantly computed data, we record the redundant times and then average the accumulated sum for correctness.

We next discuss the number of rows $N$. Without loss of generality, here we use BP as an example to infer $N\!=\!N_{\mathrm{BP}}$. We can extend it to $N_{\mathrm{FP}}$ to infer a customized value for FP. Before detailed analysis, we first need to analyze the volume of overlapped/replicated data.

We employ a bottom-top manner to infer the volume of overlapped data $o_{r}^0$ in the input matrix, i.e., from the $L$-th layer to the zero/input layer. $o_{r}^0$ is determined in a recursive manner, as shown in Eq.~(\ref{Equation:overlapped_size}), where $b^l$ indicates the volume of data required by $z_r^l$ but physically partitioned into $z_{r+1}^l$.

\begin{equation}
\label{Equation:overlapped_size}
~\left\{
\begin{aligned}
&o_{r}^{0} = (o_{r}^{1} - 1) \cdot s^1+ k^1 \\
&o_{r}^l = (o_{r}^{l+1}  -1 ) \cdot s^{l+1} + k^{l+1} \\
&o_{r}^{L-1} = b^{L - 1}
\end{aligned}
\right.
\end{equation}

Similarly, we can calculate $o_{r+1}^0$. For a balanced partitioning, we enforce $o_{r+1}^{0} = o_{r}^{0}$. To achieve this goal, we should guarantee that $a^{l}$ is equal to, or at least approximates $b^{l}$, as much as possible. Here $a^l$ indicates the volume of data required by $z_{r+1}^l$ but physically partitioned into $z_r^l$. Then the total size of overlapped data between two consecutive rows at the $l$-th layer is $o^{l} =o_{r}^{l} + o_{r+1}^{l}$.

Furthermore, to optimize GPU memory usage, it is crucial to balance the GPU memory space required for both forward and backward propagation computations in each row. Initially, we evenly partition the input data $\chi$ into $N$ rows, resulting in ($N-1$) portions of replicated data. For each $\chi_{r}$, where $r \in [2, N-1]$, its top boundary and bottom boundary both require the whole overlapped/replicated data respectively from $\chi_{r-1}$ and $\chi_{r+1}$. However, $\chi_{1}$ and $\chi_{N}$ only need completely overlapped data from one side. Therefore, for $\chi_{r}$, where $r \in [1, N]$, if the sizes of completely overlapped data obtained from each side are the same, then the size of $\chi_{r}$ as a middle row will be larger than those of $\chi_{r-1}$ and $\chi_{r+1}$. That clearly destroys the balance of memory consumption across rows. Therefore, the ranges of partitioning for $R_1$, $R_N$, and middle $R_r$, need to be adjusted according to the actual situation, to ensure that the total size of each original row and its assigned overlapping data is as similar as possible. An effective approach is to evenly divide the last layer, and then use the row size to calculate the total size of each row of the original input data and overlapped data after deconvolution. For the $R_1$ and $R_N$, by subtracting the size of overlapped data, we can immediately know the size of the original partitioned data. For other rows $R_i$, $i\in [2, N-1]$, we need to perform subtraction twice to correctly infer the size of the original partitioned data.

Now we can compute the number of rows $N$ used in BP. We still follow the two principles introduced in Sec.~\ref{section:3row:backward}. 
$N$ can be determined by the following inequalities,
\begin{align*}
\min(N) \quad &\text{s.t.} \\
&\frac{\sum_{l=1}^{L-1}\varrho^{l} + B(N-1)\sum_{l=1}^{L}o^{l}}{N} + \varrho^{L} + \xi < \mathcal{M}.
\end{align*}

\stitle{Time and Space Complexities.}
In the overlapping scheme, the space usage for each row is balanced. We then report the complexity of each row as below.
$$\frac{\sum_{l=1}^{L-1}\varrho^{l} + B(N-1)\sum_{l=1}^{L}o^{l}}{N} + \varrho^{L}$$

We next analyze the time complexity. For FP, it consists of the costs of processing the original partitioned data and the overlapped data. Here we only care about the cost during convolutional layers and exclude the cost of full connection networks, since our techniques only work for the former.

For the original data, the time complexity is given by
$$
\tau = \sum_{l=1}^{L} 2 \cdot (k^{l})^2 \cdot B \cdot C^{l-1} \cdot C^{l} \cdot H^l \cdot W^l.
$$

For overlapped data, it is expressed as
$$
\iota = \sum_{l=1}^{L} 2 \cdot (k^{l})^2 \cdot B \cdot (N-1) \cdot C^{l-1} \cdot C^{l} \cdot o^{l} \cdot W^l.
$$

Since two FPs and one BP are required in convolutional layers, the total time complexity for the overlapping scheme is then $4(\tau + \iota)$.

The overlapping scheme introduces an additional forward propagation calculation for the original complete data in one convolutional layer, as well as two forward propagation calculations and one backward propagation calculation for the overlapped data, compared to traditional CNNs. Therefore, the additional time complexity of the overlapping scheme compared to the traditional column-wise CNN is $\tau + 4\iota$.

Below, we also report the time complexity of \emph{2PS}. Note that now there is no redundant computation. Therefore, its computational complexities for both FP and BP are the same as that of traditional column-wise CNNs. However, in BP, an additional FP calculation is still performed to recover disappeared feature maps released in FP. Thus, the time complexity of \emph{2PS} can be represented using the time complexity of traditional column-wise CNNs. It is $4\tau$. Compared to column-wise CNNs, our \emph{2PS} has an additional term $\tau$. Compared to the overlapping scheme, it reduces the time complexity by $4\iota$.


\stitle{Impact of $N$.}
According to Eq~(\ref{Equation:overlapped_size}), the overlapped size $o_r^0$ is proportional to the number of convolutional layers $L$. Assuming we uniformly partition the data into $N$ rows, the height of the $r$-th row in the partitioned original input is $H_r^0 = H/N$. $N$ is inversely proportional to $H_r^0$. However, $o_r^0$ is solely related to the number of layers involved in partitioning. Thus, when $H_r^0 < o_r^0$ (i.e., $N > H/o_r^0$), it becomes impossible to obtain sufficient overlapped data, rendering this scheme ineffective. But sometimes we still need to increase $N$ while maintaining $N \leq H/o_r$, for better memory reduction. Then the key is to reduce the value of $o_r^0$. Achieving this reduction only requires to minimize the number of layers involved in partitioning.

Therefore, we introduce the \emph{OverL-H} scheme, which integrates the checkpoint method. By strategically placing checkpoints in the network layer and performing partitioning between any two checkpoints, we can increase the allowable value of $N$ between checkpoints. However, this flexibility comes at the expense of sacrificing the spatial occupation of the layer where the checkpoint is located. Thus, in practical applications, a comprehensive consideration is necessary to determine the optimal checkpoint locations. Fortunately, this approach has been studied in Ref.~\cite{reference:34}.


\section{EXPERIMENTS}\label{section:5exp}
Now we study the performance of our proposals by comparing them with state-of-the-art competitors.

\subsection{Experimental Setups}\label{section:5exp:setting}
Before detailed analysis, we first list the overall settings in experiments for better understanding.

{\stitle{Hardwares, Benchmarks and Datasets.}}
We run experiments on two GPU servers. One is Dell Precision equipped with NVIDIA GeForce RTX 3090 GPU, 10496 CUDA Cores @1.70GHz, 24GB HBM2; and two Intel XEON Silver 4216 Processors, 64GB DDR4 RAM. The other is LENOVO with NVIDIA GeForce RTX 3080 GPU, 8704 CUDA Cores @1.71GHz, 10GB HBM2; and Intel(R) Core(TM) i9-10900K CPU, 64GB DDR4 RAM. Both servers run Ubuntu 20.04.1 LTS, CUDA Toolkit 11.6, cuDNN 8.0.5, and transfer data via PCIe 3.0. We use two representative CNN models VGG-16 and ResNet-50 as benchmarks. They have different architectures and hence can validate the adaptability of our proposals. They are trained with a subset of the ImageNet dataset~\cite{imagenet}. It consists of 13,000 images labeled with one of 10 exclusive classes. By default, the dimension of each image is conventionally resized to $H\!=\!224\!\times\!W\!=\!224\!\times\!3$\footnote{Each RGB image has three color channels.}.

{\stitle{Compared Solutions.}} Our two proposals with OverLapping (Sec.~\ref{section:4opt:overlap}) and Two-Phase Sharing (Sec.~\ref{section:4opt:sharing}) designs are denoted by \emph{OverL} and \emph{2PS}, respectively. Other competing solutions include Checkpointing (\emph{Ckp}) and Offloading (\emph{OffLoad}). \emph{Ckp} optimizes memory by archiving partial data in FP and recomputing others in BP~\cite{reference:34}. Authors give a preferred checkpointing frequency and location selection guide to strike a good space balance between archiving and recomputing operations, to maximize the memory reduction. On the other hand, \emph{OffLoad} offloads data from GPU to CPU and enhances the efficiency by overlapping tensor computation and data migration~\cite{reference:28,reference:30}. Its newly released open-source implementation further allows to control the volume of offloaded data in a fine-grained manner~\cite{reference:32}. We then carefully select the best ratio via multiple attempts, to fully utilize the aggregated capacity of GPU HBM2 and CPU RAM. All of these solutions are implemented on the widely used platform PyTorch, to perform an end-to-end fair comparison. In particular, we implement our proposals and \emph{Ckp} by own, and directly run the released version for \emph{OffLoad}~\cite{reference:32}. We also include the recently published \emph{Tsplit}~\cite{reference:36} solution, which achieves the best memory reduction as reported, by splitting tensors and combining the ideas of \emph{Ckp} and \emph{OffLoad}. Our solutions also employ the idea of abandoning cheap-to-be-recomputed feature maps used in \emph{Tsplit}~\cite{reference:35,reference:36}. Note that its PyTorch implementation is not open-sourced, but we can directly quote some key reported figures, since it runs tests on a Titan RTX GPU, which has the same memory as our GeForce RTX 3090 device. Besides, we combine our proposals with \emph{Ckp} for better memory scalability. We use a suffix ``H'' to stand for the hybrid variants, namely \emph{OverL-H} and \emph{2PS-H}. Finally, to show advantages of solutions above, we create \emph{Base} by running the original PyTorch implementations.

\subsection{Overall Performance Comparison}\label{section:5exp:overall}
We first study the overall performance of all solutions, in terms of memory reduction and runtime latency.

{\stitle{Memory Reduction.}} Given a model architecture, there exist two factors dominating the memory footprint, i.e., the batch size and the image dimension. We thereby measure the memory scalability by respectively reporting the largest values of each factor that every solution can reach. On other hand, VGG-16 and ResNet-50 have different layer numbers, that explores the impact of network depth. In particular, given an image with $H\!\times\!W\!\times\!3$ and $H\!=\!W$, we report $H$ as its dimension metric, for simplicity. When exploring the scalability, we expand $H$ and $W$ with equal proportion by concatenating partial or the whole of other original images.

\begin{figure}[htbp]
	\centering
      \includegraphics[width=0.6\linewidth]{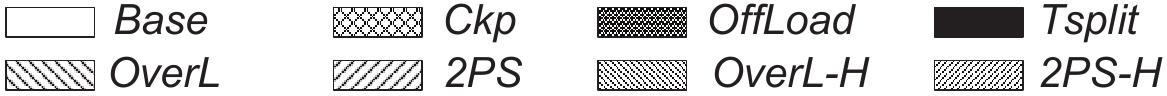}
    \renewcommand{\thesubfigure}{\scriptsize (\alph{subfigure})\space}
	\subfigure[RTX3090-(24+64)GB]{
			 \includegraphics[width=0.5\linewidth]{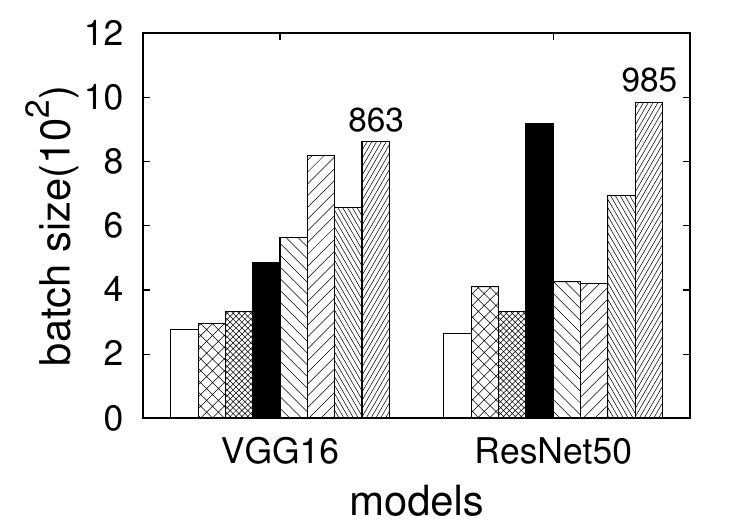}
	}%
	\subfigure[RTX3080-(10+64)GB]{
			 \includegraphics[width=0.5\linewidth]{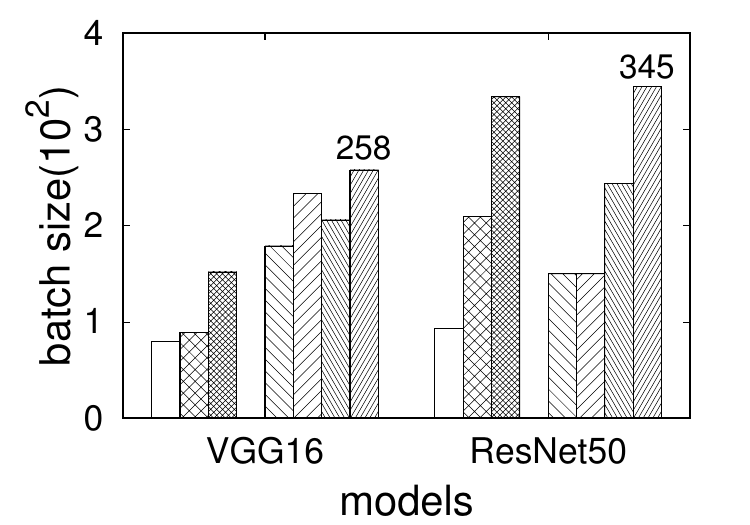}
	}
	\centering
	\caption{The largest batch size comparison}
	\vspace{0.2cm}
	\label{figure:batchsize}
\end{figure}

\begin{figure}[htbp]
	\centering
    \includegraphics[width=0.6\linewidth]{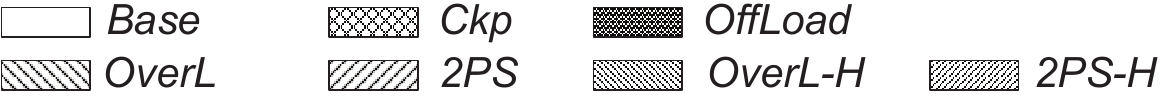}
    \renewcommand{\thesubfigure}{\scriptsize (\alph{subfigure})\space}
	\subfigure[RTX3090-(24+64)GB]{
			 \includegraphics[width=0.5\linewidth]{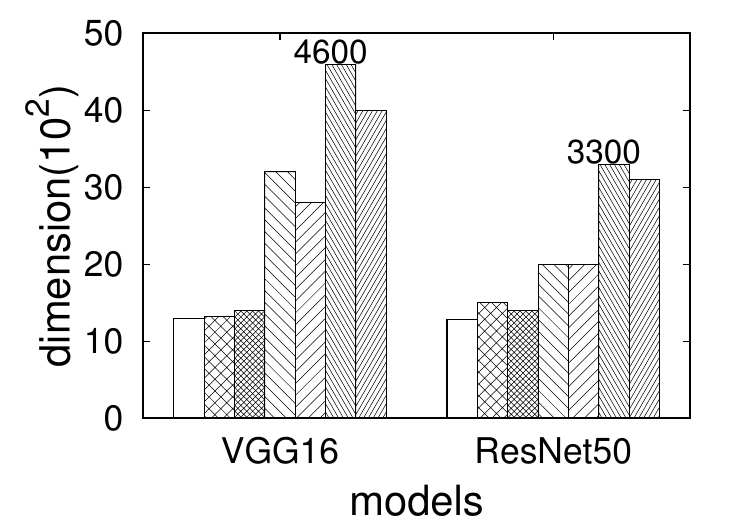}
	}%
	\subfigure[RTX3080-(10+64)GB]{
			 \includegraphics[width=0.5\linewidth]{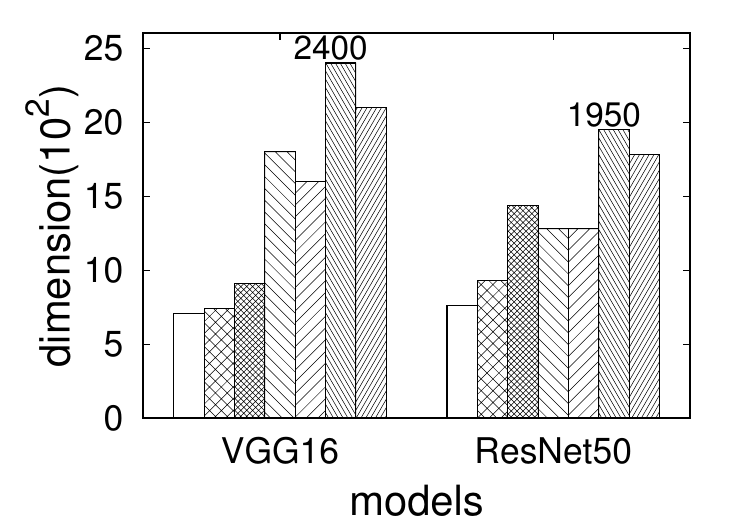}
	}
	\centering
	\caption{The largest image dimension comparison (batch size 8)}
	\vspace{0.2cm}
	\label{figure:resolution}
\end{figure}

Figs.~\ref{figure:batchsize}-\ref{figure:resolution}\footnote{For \emph{Tsplit}, in the original literature, the image dimension factor is not tested, and also, no device with the same 10GB HBM2 like our RTX3080 is used. Thus, it is missing in some figures.} depict evaluation results. In summary, \emph{Offload} outperforms \emph{Ckp} since the former can additionally utilize CPU memory resources. Nevertheless, our basic \emph{OverL} and \emph{2PS} still beat \emph{Offload} due to reusing memory across rows. The respectively largest improvements are 69\% and 146\% for batch size; and 423\% and 300\% for image dimension. Further, benefitting from \emph{Ckp}, our hybrid variants \emph{OverL-H} and \emph{2PS-H} can perform much deeper and more fine-grained row partitioning, to enhance memory sharing (explained later). They thereby create another scalability gap where the largest improvements increase to 108\% and 195\% for batch size; and 980\% and 716\% for image dimension. Finally, compared with the up-to-date \emph{Tsplit}, our basic \emph{OverL} and \emph{2PS} improve the batch size by 16\% and 69\%, respectively. For our hybrid variants, the factors increase to 36\% and 78\%, respectively.

Note that \emph{OverL} and its variant underperform the corresponding two-phase-sharing counterparts. This is because here we try our best to increase the number of rows to maximize memory reduction; then the rapidly growing overlapped, redundant data in the former consume more spaces than the preserved, shared data.

There is no doubt that for all solutions, their performance on RTX3090 with more memory resources is better than that on RTX3080. An in-depth comparison between subfigures (a) and (b) further reveals that the scalability gap between our proposals and \emph{OffLoad} narrows from RTX3090 to RTX3080. The explanation is that now our row-partitioning only works in GPU HBM2, which is sensitive to the resource reduction on RTX3080; while, for \emph{OffLoad}, such impact can be partially offset by utilizing the CPU RAM. This motives us to combine our proposals and the offloading technique in future.

\eat{
In particular, the channel number varies with layers. We proportionally expand these numbers but only report the number of the first layer, for simplicity.
An in-depth comparison between Fig.~\ref{figure:batchsize} and Fig.~\ref{figure:channel} reveals that, no solution can proportionally translate the batch size reduction into the channel number increase. The main reason is that a large channel number increases the memory consumption of not only feature maps, but also model parameters. Such negative impact is particularly heavy for our GPU-only proposals, especially for RTX3080, because of the aggressive contention on limited GPU memory. Exceptions are \emph{OffLoad} and \emph{Tsplit} as the large aggregated memory amortize the impact, even making them better than our proposals.
}

At last, we use the batch size test as an example to show how checkpointing enhances the performance of our designs. Our metrics are the number of layers involved in row-centric update, and the sum of rows in every such layer. For both metrics, the more, the better. Table~\ref{table:help} lists the comparison result. The significant increase explains why \emph{OverL-H} and \emph{2PS-H} outperform their original versions.

\begin{table}[htbp]
\caption{Exploring the impact of checkpointing on \emph{OverL} and \emph{2PS}.}
\begin{center}
\begin{tabular}{|l|r|r|r|r|}
\hline
\multirow{2}{*}{\textbf{Solutions}}
&\multicolumn{2}{|c|}{\textbf{VGG-16}}
&\multicolumn{2}{|c|}{\textbf{ResNet-50}}\\
\cline{2-5}
\textbf{}
&\textbf{\# of Layers} &\textbf{\# of Rows}
&\textbf{\# of Layers} &\textbf{\# of Rows}\\
\hline
\emph{OverL}& 6& 42& 10& 80\\
\emph{OverL-H}& 13& 54& 22& 128\\
\emph{2PS}& 7& 42& 10& 40\\
\emph{2PS-H}& 13& 54& 49& 142\\
\hline
\end{tabular}
\label{table:help}
\end{center}
\end{table}

\begin{figure}[htbp]
	\centering
 \includegraphics[width=0.6\linewidth]{Legend_dimension.pdf}
    \renewcommand{\thesubfigure}{\scriptsize (\alph{subfigure})\space}
	\subfigure[RTX3090-(24+64)GB]{
			 \includegraphics[width=0.48\linewidth]{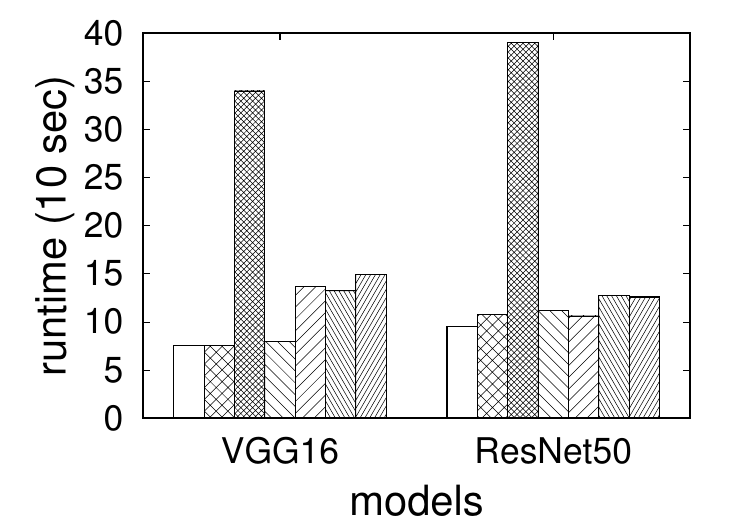}
	}%
	\subfigure[RTX3080-(10+64)GB]{
			 \includegraphics[width=0.48\linewidth]{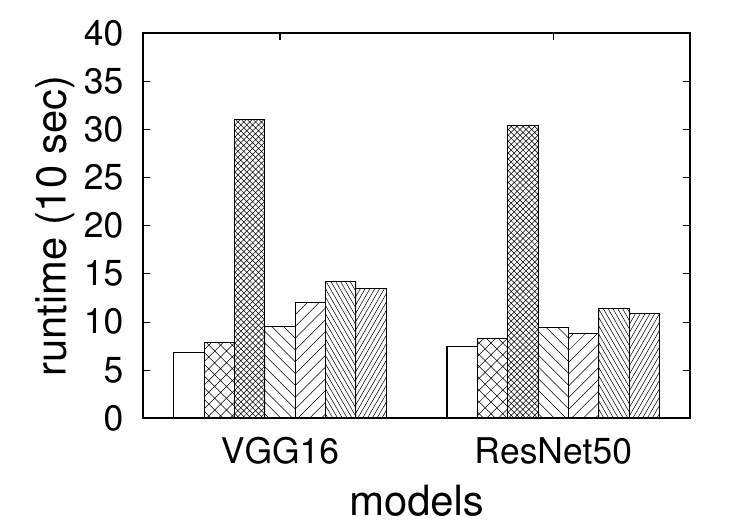}
	}
	\centering
	\caption{The runtime latency comparison using the same settings in Fig.~\ref{figure:batchsize}}
	\vspace{0.2cm}
	\label{figure:runtime}
\end{figure}

{\bf\stitle{Runtime Latency.}} Taking the batch size test as an example, Fig.~\ref{figure:runtime} reports the corresponding runtime costs per epoch. We clearly see that all solutions trade efficiency for memory scalability. Among them, \emph{OffLoad} has the most series runtime latency (up to 356\% slower than \emph{Base}). This is because transferring a lot of data between GPU and CPU poses great challenges to the limited PCIe bandwidth. \emph{Ckp} needs to recompute feature maps, yielding slight efficiency penalty (only 15\%) due to the high compute power of GPU. While, our proposals additionally pay for coordination between rows, including redundant computations for overlapped dimensions, and update interruption for sharing intermediate results. The latency factors of basic versions thereby increase to 40\% and 81\%, respectively. For hybrid versions with checkpointing, these factors are respectively 108\% and 99\%.

\subsection{Evaluation on Scalability}\label{section:5exp:scalability}
\begin{figure}[htbp]
	\centering
         \renewcommand{\thesubfigure}{\scriptsize (\alph{subfigure})\space}
	\subfigure[RTX3090-24GB]{
			 \includegraphics[width=0.48\linewidth]{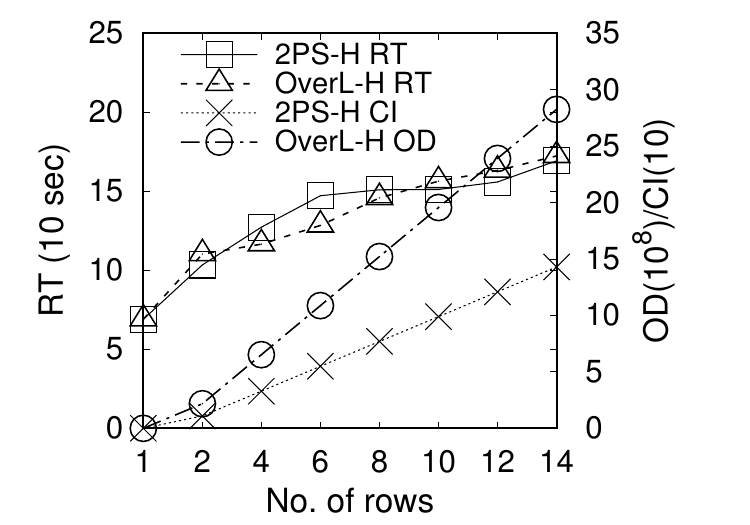}
	}%
	\subfigure[RTX3080-10GB]{
			 \includegraphics[width=0.48\linewidth]{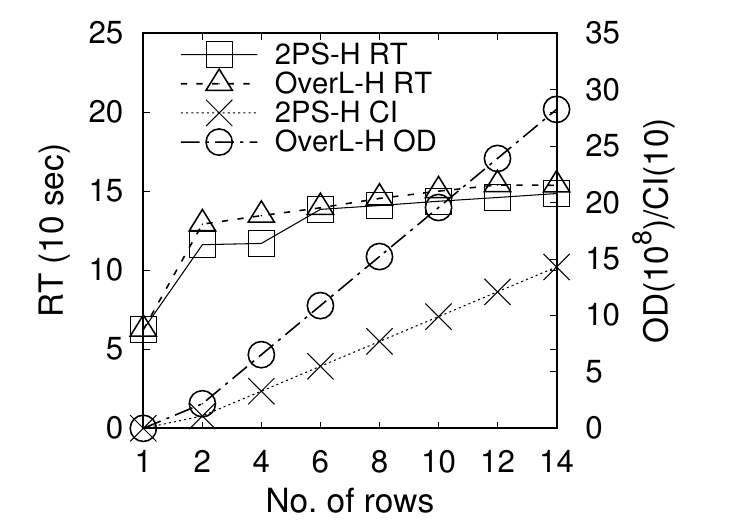}
	}
	\centering
	\caption{Training runtime (VGG-16, batch size 64)}
	\vspace{0.2cm}
	\label{figure:scaruntime}
\end{figure}

\begin{figure}[htbp]
	\centering
    \renewcommand{\thesubfigure}{\scriptsize (\alph{subfigure})\space}
	\subfigure[Overall consumption]{
			 \includegraphics[width=0.48\linewidth]{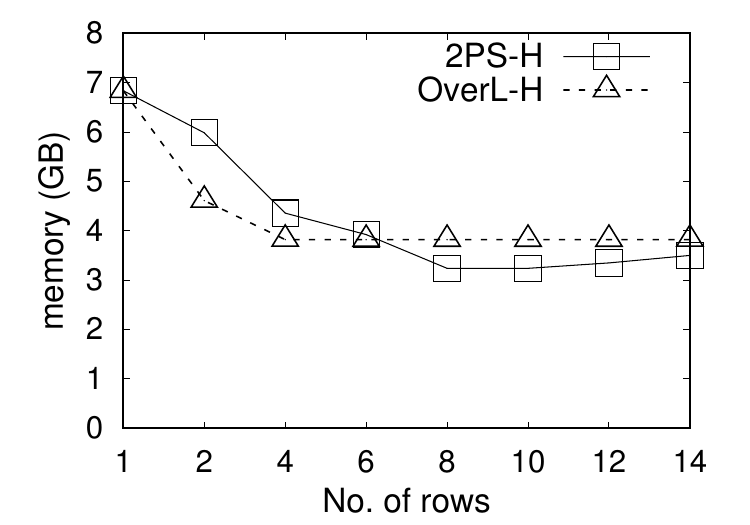}
	}%
	\subfigure[Additional consumption]{
			 \includegraphics[width=0.48\linewidth]{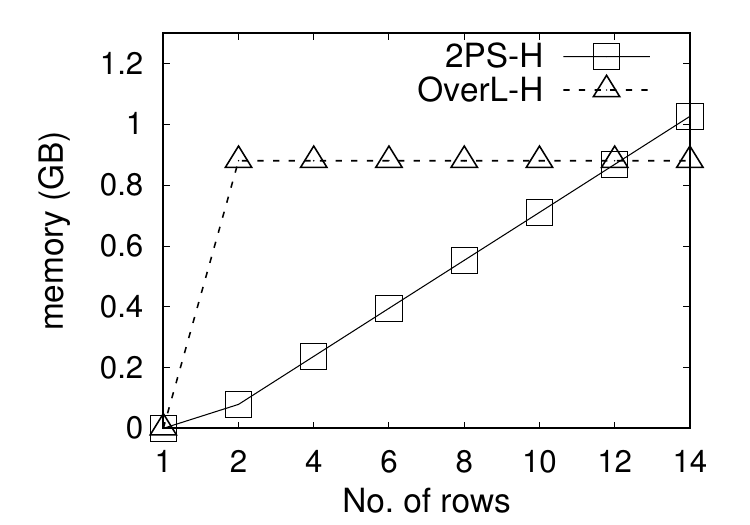}
	}
	\centering
	\caption{Memory consumption (VGG-16, batch size 64, RTX3090)}
	\vspace{0.2cm}
	\label{figure:scamemory}
\end{figure}

As an indication of how our proposals scale, Figs.~\ref{figure:scaruntime} and \ref{figure:scamemory} show the impact of row partitioning granularity $N$ on training runtime and memory consumption. Previous experiments have validated that \emph{OverL-H} and \emph{2PS-H} achieve the maximum input scale in terms of batch size and image dimension. Now we only test them for brevity. We manually increase $N$ from the original 1 (the default PyTorch) to 14, and then observe their different behaviors.

Fig.~\ref{figure:scaruntime} depicts a sublinear RunTime (RT) increase for both approaches. As explained in Fig.~\ref{figure:runtime}, such variation is mainly caused by the coordination operation. Now we quantify it for better understanding. In \emph{OverL-H}, the cost is directly proportional to the total number of Overlapped Dimensions (OD) among all rows. While, in \emph{2PS-H}, it can be measured by how many times the normal Computation Interruption (CI). We report the two counters against granularity. As expected, both of them exhibit linear increase. Another observation is that the runtime comparison result of our approaches is very different on the two devices. We give our explanation below. For \emph{OverL-H}, its redundant computations caused by overlapped dimensions are easy-to-be-parallelized. Fortunately, RTX3090 can offer enough power for the parallel acceleration, and hence the de facto runtime latency is less significant. By contrast, the interruption of \emph{2PS-H} is insensitive to the power. Thus, it can beat \emph{OverL-H} on the lower-configured RTX3080.

With the same settings, Fig.~\ref{figure:scamemory}(a) reports overall memory consumption versus granularity $N$. Both \emph{OverL-H} and \emph{2PS-H} can effectively reduce the resource requirement. They achieve the best performance roughly at $N\!=\!8$, yielding up to 44\% and 53\% reductions, respectively. Note that the slope of the reduction curve gradually becomes gentle for the two approaches, and even reverses for \emph{2PS-H}. As shown in Fig.~\ref{figure:scamemory}(b), for \emph{2PS-H}, this is mainly due to the growing volume of Sharing Data (SD) accumulated across all rows (used in FP and BP), that possibly offsets the reduction effort. While, for \emph{OverL-H}, the volume is related to the number of layers, rather than row granularity.This is because the independent advantage allows to safely remove data associated with a row, once its processing is completed. However, with increase of $N$, the Overlapped Data (OD) will dominate the memory consumption, weakening the benefit of our row-centric design. These phenomenons experimentally validate our informal analysis about $N$ in Secs.~\ref{section:4opt:sharing} and \ref{section:4opt:overlap}. Such difference also leads a turning point of performance comparison around $N\!=\!6$. Together with the conclusion draw in Fig.~\ref{figure:scamemory}, we state that \emph{OverL-H} is a prominent solution if enough memory and compute resources can be provided; otherwise, \emph{2PS-H} is a preferred alternative.

\subsection{Validation of Convergence}\label{section:5exp:convergence}
\begin{figure}[htbp]
	\centering
         \renewcommand{\thesubfigure}{\scriptsize (\alph{subfigure})\space}
	\subfigure[VGG-16]{
			 \includegraphics[width=0.48\linewidth]{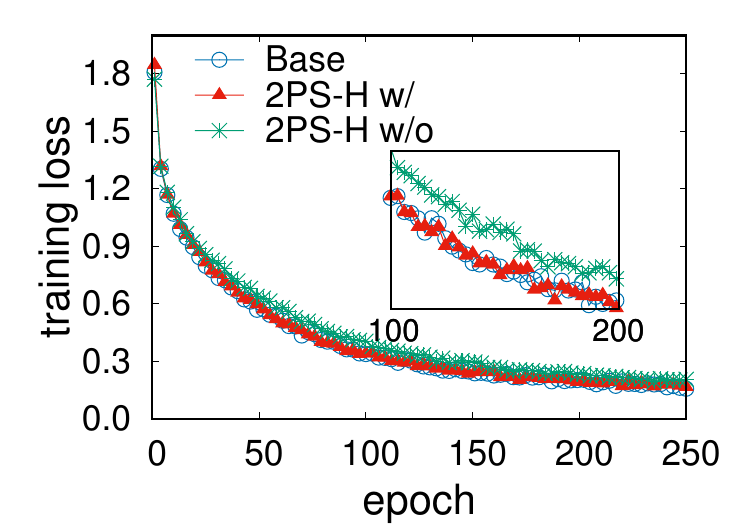}
	}%
	\subfigure[ResNet-50]{
			 \includegraphics[width=0.48\linewidth]{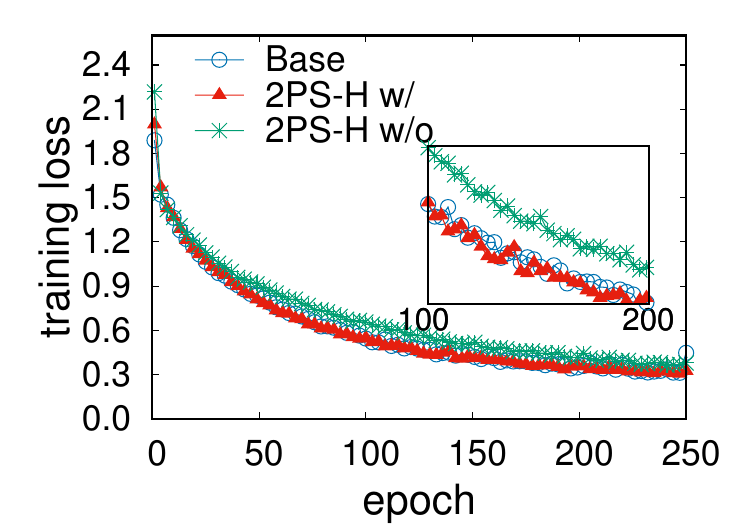}
	}
	\centering
	\caption{Convergence analysis (RTX3090, batch size 256)}
	\vspace{0.2cm}
	\label{figure:correct}
\end{figure}

We finally validate the necessary of inter-row coordination, by keeping track of the training loss as epochs proceed. We select \emph{2PS-H} as an example to create two branches w/ and w/o data sharing, and compare them against \emph{Base}. Fig.~\ref{figure:correct} shows loss vs. epoch plots for VGG-16 and ResNet-50. We can find \emph{2PS-H} w/ sharing performs very similarly with \emph{Base}. However, the abnormal version has a big penalty especially in late epochs where the accumulated errors possibly force training to make a long detour to converge.

\section{Related Works}\label{section:6related}
This section outlines existing efforts for memory scalability. Although they are layer/column-based and hence orthogonal to our row-centric idea, we still benefit from some prominent designs like abandoning cheap-to-be-recomputed activation data~\cite{reference:35,reference:36} and checkpointing~\cite{reference:34}. Below we overview them from external auxiliary and internal modification aspects.

\subsection{External Auxiliary with Additional Hardware Requirements}\label{section:6related:external}
The idea behind external auxiliary techniques is to aggregate resources of additional hardwares. The increased memory can accommodate more data, but at expense of economic investments. Our techniques can be integrated with them to reduce such costs. These techniques fall into two categories, parallel computation and offloading data.

\stitle{Parallel Deep Learning.}
The most widely used parallel training policy is data parallelism (DP) where the input samples are distributed across devices but parameters are fully replicated everywhere~\cite{dataparallel,reference:1}. It achieves prominent batch size scalability, since the corresponding feature maps are also distributed with samples. Some early works also follow the DP design but split every single sample matrix across devices, to perform training or inference on low-configured mobile hardwares~\cite{reference:8,reference:9}. Although their partitioning policy is similar with ours but the propagation is still layer-centric, without any memory reduction. Split-CNN~\cite{reference:37} performs a row-centric scheduling but ignores the dependency between rows. The accuracy loss proportionally increases with the number of rows, leading to poor scalability. Note that DP cannot work once the model size exceeds the capacity of a single device. That can be solved by tensor parallelism~\cite{reference:2,reference:3,reference:4,reference:7} and pipeline parallelism~\cite{reference:5,reference:6} where parameters are split in vertical and horizontal directions, respectively, but samples are the same for all devices. Now feature maps are distributed with parameters.

\stitle{Offloading Data from GPU to CPU.}
Recall that the modern accelerator like GPU cannot work independently. Its accompanied CPU can read/write data in the much bigger RAM. That motivates researchers to offload data, like feature maps and parameters, from GPU to CPU, to make full use of available resources~\cite{reference:28,reference:29}. The capacity of available memory indeed increases, but frequently transferring data between CPU and GPU incurs expensive runtime latency. ZeRO-Offload~\cite{reference:30} optimizes the latency by overlapping computation and communication, and offloading parameter update to CPU to reduce the frequency of data transmission. While, CPU possibly becomes the new bottleneck since ZeRO-Offload cannot flexibly control how many data and workloads are offloaded. ZeRO-infinity~\cite{reference:31} even utilizes the out-of-core storage, which severely degrades the training speed. The recently released open-source platform Hfai~\cite{reference:32} inherits the overlapping design and implements the fine-grained control for offloading. But the latency is still large based on our tests. Another offloading variant, SuperNeurons, permanently and completely removes cheap activations and offloads others~\cite{reference:35}. That can use the aggregated memory to accommodate larger models and reduce runtime latency. We borrow this idea to exclude cheap functions from our row-centric change for efficiency. Tsplit~\cite{reference:36} also follows it and further devises a model-guided approach to enhance scalability.

\subsection{Internal Modification with Potential Accuracy Loss}\label{section:6related:internal}
Different from hardware-dependent external techniques, another economically feasible line is to modify the internal data management policy. The representative techniques include checkpointing, and compressing data and networks. Checkpointing~\cite{reference:34} as a classic solution has already been integrated into mainstream deep learning platforms, like TensorFlow and PyTorch. It periodically makes checkpoints among all $n$ layers. Feature maps except those in the checkpoint are released for memory reduction, and recomputed from the last checkpoint during backward propagation. A preferred checkpointing frequency is $\sqrt{n}$, which can well balance the costs of storing checkpoints and benefits of releasing data. Below we summarize the other two compressing works.

\stitle{Compressing Data by Quantization.}
Early pioneers aggressively express numerical variables by a uniformly shared value within a group~\cite{reference:10}, or even binary/ternary enumeration type~\cite{reference:11,reference:12}. Such coarse-grained extreme quantization generates huge errors. Recent works quantize feature maps and gradients by lower precision with fewer bit width~\cite{reference:13,reference:14,reference:15}, typically like replacing FP32 with INT8. By filtering outliers~\cite{reference:16} and modifying the model structure~\cite{reference:15}, the accuracy loss can be further reduced. ZeRO++~\cite{reference:17} uses a fine-grained blocking mechanism for high-level quantization, but at expense of runtime penalty. Besides, there are also lossless approaches by distinguishing data features. For example, the output of the ReLU function can be expressed with only 1 bit~\cite{reference:18}. Also, it is enough to represent activations with FP16 but for fine-tuned parameters, FP32 is preferred~\cite{reference:19}.

\stitle{Trimming Network Structure.}
Some researchers remove unimportant connection weights and/or nearly-zero activations~\cite{reference:20,reference:21,reference:22}. Gradient accumulation~\cite{reference:33} divides a batch into several micro-batches where gradients are accumulated but feature maps are abandoned. The latter yields additional errors. Others directly focus on pruning parameters, by decoupling complex network structures~\cite{reference:23} and dynamic and progressive sparsity evaluation~\cite{reference:24}. AutoLR~\cite{reference:25} proposes the hierarchical pruning to reduce network complexity. These works can significantly compress the model size and reduce associated computational complexity, but incur the loss of accuracy and generalization ability. Also, they destroy the raw ``regularity'' feature, leading to underutilization of massive SPMD threads in modern accelerators. Joo et al. analyze accuracy problem and devise a weight compensation technique~\cite{reference:26}. In fact, most of these techniques are tailored for inference, especially on low-configured mobile devices. The pruned model usually requires to be re-trained or at least fine-tuned since the network changes. Yvinec et al.~\cite{reference:27} proposes a data-independent structured pruning method without such requirement.

In this paper, we have combined checkpointing with our proposals to offer better scalability. Since our method is lossless, it can also work with quantization and network trimming techniques to alleviate their built-in drawbacks.

\section{Conclusion and Future Works}\label{section:7con}
Since CNN is developed, its training always suffers from expensive memory requirements, especially for the large volume of activations across layers. This paper mitigates such bottleneck head on by challenging the conventional wisdom of layer-centric update. The proposed row-centric optimization allows to train CNN in a fine-grained manner throughout layers, to safely reuse limited memory. Experiments validate that our two branches with very different implementations feature contributions respectively in high- and low-configured environments. They can achieve the maximum scale in most cases without additional resources, like CPU RAM.

Our row-centric design is orthogonal to and compatible with existing works. We have already integrated checkpointing into our proposals to show that such enhancement can further reduce memory consumption. However, many other efforts like offloading and splitting sensors can be explored. Besides, activations exist in a majority of deep learning networks, since they also have the multi-layer architecture and employ the FP and BP training policy. Theoretically, our proposals can be generalized to them. Building a generic framework integrating mainstream techniques is among our ongoing investigations.

\section*{Acknowledgments}
This work was supported by the National Natural Science Foundation of China ( U22A2068 and 61902366), the Fundamental Research Funds for the Central Universities (202042008), the National Key Research and Development Program of China (No. 2021YFF0704000), the National Natural Science Foundation of China (61902365 and 62072083), the National Science Foundation Grants (CNS-1815412 and CNS-1908536), and the Graduate Professional Development Fund Project of Computer Department of Ocean University of China (CSZS2022004).
\bibliographystyle{IEEEtran}
\bibliography{reference}

\end{document}